# Perspective: Speculative role of Tmp21 mediated protein secretory pathway during endoplasmic reticulum (ER) stress induced chronic inflammation

Katte Rao Toppaldoddi[1,2,3]

[1]INSERM, UMR1170, Institut Gustave Roussy, Villejuif, France, [2]Hôpital Saint-Louis, Institut Universitaire d'Hématologie, Université Paris Diderot, Paris, France and [3]St. John's Medical Hospital and College, Bangalore, India.

Correspondence (E-mail): katte.rao@gmail.com, katte-rao.toppaldoddi@inserm.fr

**By deploying myelofibrosis as the disease context, I wish to propose that increased availability of Tmp21 (an NFAT gene target[1]) induces aberrant protein secretion from the ER contributing to pathological consequences, which has not been elucidated before. Primary myelofibrosis is now mainly considered as an advanced stage of BCR-ABL1 negative myeloproliferative neoplasms (MPN), which otherwise include polycythemia vera and essential thrombocythemia. Myelofibrosis is defined by an increased insoluble collagen fiber deposition in the bone marrow[2] and harbors chronic inflammation as an important component in disease progression[2].**

It has been demonstrated that the event responsible for MPN is the uncontrolled proliferation of myeloid precursors in the bone marrow due to somatic mutations that constitutively activate the JAK-STAT pathway through cytokine receptors leading to cell transformation[3-8]. In primary myelofibrosis, these mutations are associated with other mutations in several pathways, which may induce differentiation defects (myelodysplastic features). The key cell type in the development of this fibrosis appears to be megakaryocytes which secrete numerous pro-inflammatory protein molecules in an uncontrolled manner, causing aberrant re-programming of bone marrow mesenchymal stromal cells (BM-MSCs), extracellular matrix proteins secretion and inhibition of normal hematopoiesis[2].

In myelofibrosis associated megakaryocytes, upregulation of NFAT gene expression via Flt3 hyperactivation has been reported in the context of inflammation[2] and Tmp21 is one of the NFAT targets expressed in majority of the human tissues[9], which may be involved in this abnormal protein secretion.

First, Tmp21 is an ER-Golgi associated protein cargo receptor which gets substantially activated to secrete proteins only upon ER stress to inhibit development of acute ER stress by establishing adaptive ER homeostasis. Further, ER resident misfolded GPI anchored proteins access extracellular space transiently through ER stress induced activation of Tmp21 mediated secretory pathway. Moreover, during ER stress conditions Tmp21 dominates over Calnexin function (soluble/mature protein fold inducing ER chaperone) and secretes insoluble/misfolded proteins to extracellular space[10] and in addition, increased availability of Tmp21 induces secretion of immature amyloid precursor protein (APP) by inhibiting maturation of nascent APP polypeptide chain[11], which altogether would hint that increased activation of Tmp21 mediated secretory pathway could have detrimental effects on the secretory proteins concerning their normal protein quality assessment and regulated secretion. Second, inhibition of Tmp21 causes apoptosis of transformed cells[12,13], which indicates that Tmp21 mediated protein secretory pathway plays an active role in survival and proliferation of transformed cells.

Based on the so far mentioned experimentally proven facts, I wish to put forward the model for the responsible mechanism by explaining it in myelofibrosis associated megakaryocytes as constitutive activation of STAT-5 and STAT-3 transcriptional factors cause over-expression of secretory proteins including pro-inflammatory molecules[14] resulting in ER protein folding overburden that may naturally lead to ER stress. Thus, ER stress activates Tmp21 mediated secretory pathway which may cause aberrant secretion of immature/insoluble pro-inflammatory molecules to the extracellular space leading to aberrant and transient activation of pro-inflammatory pathways causing initiation of chronic inflammation and pathological re-programming of BM-MSCs. Therefore, detailed investigation of the effects of Tmp21 inhibition as a novel targeted drug therapy may provide immediate clinical benefits in the treatment of myelofibrosis associated chronic inflammation. Moreover, as ER stress induced chronic inflammation persists in several different pathological conditions and are difficult to treat with the currently available treatment regimens[15], hence, Tmp21 inhibitors may serve as targeted drug therapy to treat such diseases. Ultimately, Tmp21 inhibitors development may replace the therapeutic strategy of directly targeting ER stress activated unfolded protein response (UPR) pathway to avoid adverse side effects of UPR inhibitors.

Acknowledgements:
I am thankful to my PhD supervisor Dr. William Vainchenker at Institut Gustave Roussy, for the critical discussions. I am supported by fellowships from L'Institut national de la santé et de la recherche médicale (INSERM), France, GLUE Grant Initiative, Department of Biotechnology, Governement of India and travel scholarship from the Service for Science and Technology at the French Embassy (SST), New Delhi, India.